\documentclass[%
reprint,
amsmath,assume,
aps,
prb,
]{revtex4-2}

\usepackage{graphicx}% Include Figure files
\usepackage{color}
\usepackage{hyperref}

\usepackage{dcolumn}% Align table columns on the decimal point
\usepackage{bm}% bold math
\usepackage{here}
\usepackage{amsmath}
\usepackage{ulem}
\usepackage[utf8]{inputenc}
\usepackage[T1,T2A]{fontenc}
\DeclareUnicodeCharacter{2212}{-}
\newcommand{\add}[1]{\textcolor{black}{#1}}	%←追加
\newcommand{\Tczero}{$T_{\mathrm{c}}^{\mathrm{zero}}$ }
\newcommand{\Tconset}{$T_{\mathrm{c}}^{\mathrm{onset}}$ }

\newcommand{\Tc}{$T_{\mathrm{c}}$ }

\begin{document}

%\preprint{APS/123-QED}

\title{Study on fluctuations of interface-enhanced superconductivity in ultrathin FeSe/SrTiO$_3$ by the Nernst effect}% Force line breaks with \\
%\thanks{A footnote to the article title}%

\author{Tomoki Kobayashi, Ryo Ogawa, and Atsutaka Maeda}
 %\altaffiliation[Also at ]{Physics Department, XYZ University.}%Lines break automatically or can be forced with \\
%\author{Second Author}%
 %\email{Second.Author@institution.edu}
\affiliation{%
 Department of Basic Science, The University of Tokyo, 3-8-1 Komaba, Meguro-ku, Tokyo 153-8902, Japan
}%

%\author{Charlie Author}
 %\homepage{http://www.Second.institution.edu/~Charlie.Author}
%\affiliation{
 %Second institution and/or address\\
 %This line break forced% with \\
%}%
%\affiliation{
% Third institution, the second for Charlie Author
%}%

\date{\today}% It is always \today, today,
             %  but any date may be explicitly specified

\begin{abstract}
Ultrathin FeSe films on SrTiO$_3$ substrate show interface-enhanced superconductivity.
However, how the superconductivity is established including superconducting fluctuations remains unclear.
This study investigates the Nernst effect, which is sensitive to superconducting fluctuations, in ultrathin FeSe films on SrTiO$_3$.
Temperature dependence of Nernst signals in the normal state is similar to bulk FeSe, suggesting that the electrons of SrTiO$_3$ are transferred only to a few layers near the FeSe/SrTiO$_3$ interface.
\add{The Nernst effect caused by SC fluctuations was observed only below $T\sim1.2\ T_\mathrm{c}^{\mathrm{onset}}$ within our measurement resolution},
which is similar to other Fe chalcogenide systems.
Our results suggest that the pseudogap in monolayer FeSe/STO possibly originates in other electronic states rather than superconductivity.
%This work provides valuable insights into the origin of the pseudogap and superconducting fluctuations in monolayer FeSe/SrTiO$_3$.
%\begin{description}
%\item[Usage]
%Secondary publications and information retrieval purposes.
%\item[Structure]
%You may use the \texttt{description} environment to structure your abstract;
%use the optional argument of the \verb+\item+ command to give the category of each item. 
%\end{description}
\end{abstract}

%\keywords{Suggested keywords}%Use showkeys class option if keyword
                              %display desired
\maketitle

%\tableofcontents

%\section{\label{sec:level1}Introduction:\protect\\ The line
%break was forced \lowercase{via} \textbackslash\textbackslash}

%%%%%%%%%%%%%%%%%%%%%%%%%%%
\section{Introduction}
Ultrathin FeSe/SrTiO$_3$ (STO) has attracted considerable attention owing to its higher superconducting transition temperature ($T_\mathrm{c}$) compared to that of its bulk form\cite{Wang_2012,He_2013}.
The FeSe/STO interface is repotedly plays a crucial role in increasing the $T_{\mathrm{c}}$. Charge transfer from STO\cite{Liu_2012,Zhang_2017,Zhao_2018} and possible electron-phonon coupling between electrons in FeSe and phonons of STO\cite{Lee_2014,Zhang_2016_Kdose,Song_2019} have been studied as possible mechanisms causing \Tc enhancement.
%The \Tc enhancement was also observed in multilayer films by transport measurements, however, superconductivity is believed to be confined to the FeSe/STO interface because the superconducting gap was not observed on the second layer of FeSe by spectroscopic tunneling measurement(STM)\cite{Wang_2012}.
While the ultrathin FeSe/STO had been fabricated only by molecular beam epitaxy (MBE), a similar superconductivity was recently achieved by pulsed laser deposition (PLD)\cite{Kobayashi_2022,Zhao_2024}.
Similar to MBE-grown films, in PLD-grown films, the electrons in STO are transferred to FeSe, and the FeSe/STO interface is critical for achieving increased $T_{\mathrm{c}}$.
%negative hall coefficient suggesting the electron transfer from the substrate, which is consistent with MBE-grown films.
%The estimated out-of-plane coherence length $\xi_c$ was smaller than c-axis parameter of FeSe and almost independent of the total thickness of FeSe films.
Even though the thicknesses of these films are not monolayer, the measurement results of the upper critical field indicated confinement of superconductivity to a monolayer or a few layers near the FeSe/STO interface \cite{Kobayashi_2023}.

In monolayer FeSe/STO grown using MBE, the achieved \Tc was reported to be 65 K due to the gap opening at the Fermi level, as observed using angle-resolved photoemission spectroscopy (ARPES)\cite{He_2013}.
Feng group reported the establishment of superconductivity below 60–65 K, which correspond to gap-opening temperatures, using $\mu$SR\cite{Biswas_2018} and diamagnetism measurements\cite{Zhang_2015}. However, no other studies have yet reported reliable evidence of superconductivity at 65 K using transport and diamagnetism measurements.
Regardless of in-situ\cite{Pedersen_2020, Faeth_2021}or ex-situ\cite{Wang_2012} measurements, a resistive transition with onset \Tc (\Tconset) $\sim$ 40 K and zero resistivity below $T_{\mathrm{c}}^{\mathrm{zero}}<$ 29 K\cite{Wang_2012, Pedersen_2020, Faeth_2021} have been reported with an exception of a singular report\cite{Ge_2015}.
%As for the experiments for the observation of the Meissner effect, the reported values of \Tc are very scattered.
%Thus, there are questions on how these values are correlated.
%Recent studies that performed ARPES and transport measurements reported that the \Tconset was still 40 K even in the sample with the gap opening below the temperature higher than 60 K.
The observed gap at approximately 65 K may be attributed to strong superconducting (SC) fluctuations\cite{Faeth_2021, Pedersen_2020}.
However, the gap around the Fermi level is also observed in other ordered states including the pseudogap phase in cuprate\cite{Kondo_2011} and charge density wave in kagome superconductors\cite{Nakayama_2021}.
Therefore, further studies are needed to understand the relation between the gap opening temperature, $T_{\mathrm{c}}^{\mathrm{onset}}$, and $T_{\mathrm{c}}^{\mathrm{zero}}$.

One of the probes sensitive to SC fluctuations is the Nernst effect, which is the generation of a transverse electric field by a longitudinal thermal gradient.
Nernst effect was originally studied for vortex motion induced by the thermal gradient\cite{Solomon_1967,Behnia_2022}.
In the last few decades, the Nernst effect by SC fluctuations \add{has been reported.} \add{The SC fluctuations can be categorized to amplitude fluctuations, which is interpreted by short-lived cooper pairs, and phase fluctuations where proliferated vortices plays a role. The Nernst effect by amplitude fluctuatinos was reported for NbSi amorphous thin film \cite{Pourret_2006} at first. Subsequently, it was also observed in cuprate \cite{Kokanovic_2009,Chang_2012,Tafti_2014} and other amorphous or crystalline conventional superconductor\cite{Ienaga_2020,Rischau_2021}. The Nernst effect caused by phase fluctuations have been reported in hole-doped cuprate\cite{Wang_2006}, although it is still under debated\cite{Chang_2012}.}
Features of the Nernst effect caused by amplitude or phase fluctuations have been studied \add{theoretically\cite{Ussishkin_2002, Michaeli_2009, Roy_2018}}, allowing us to investigate the nature of these SC fluctuations.

In this study, we investigated the Nernst effect in ultrathin FeSe films on STO.
Temperature dependence of Nernst signals in the normal state is similar to bulk FeSe, suggesting that the electrons of STO are transferred only to a few layers near the FeSe/STO interface. \add{The Nernst effect caused by SC fluctuations was observed only below $T\sim1.2\ T_\mathrm{c}^{\mathrm{onset}}$ within our measurement resolution},
which is similar to those observed in other Fe chalcogenide systems.
The comparison to theories suggests that the observed Nernst effect caused by the SC fluctuations originates from amplitude fluctuations of a superconducting parameter.
Our results suggest that the pseudogap in monolayer FeSe/STO may originate in electronic states other than superconductivity.

%%%%%%%%%%%%%%%%%%%%%%%%%%%%%%%%%
\section{EXPERIMENTAL METHODS}
\begin{figure}[htpb]
\centering
\includegraphics[width=\linewidth]{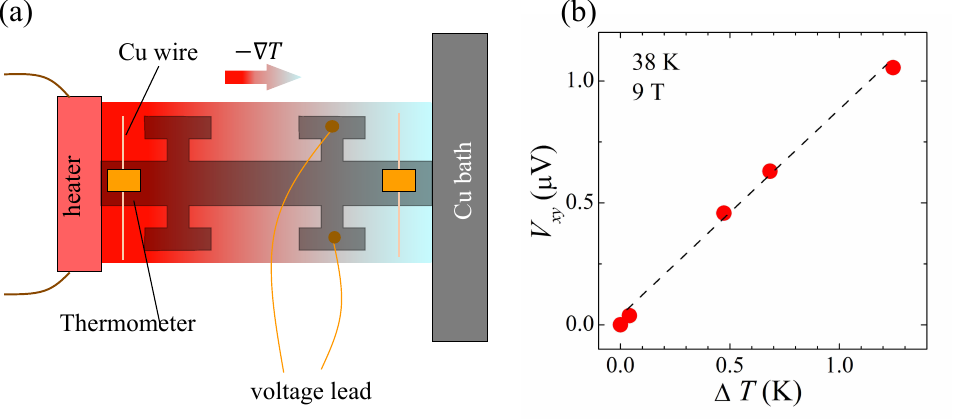}
\caption{\label{Fig:Fig1} 
(a) Schematic picture of setup used for Nernst effect measurement in this study.
(b) $V_{xy}$ as a function of $\Delta T$ at 38 K and 9 T in a 2.5 nm film.}
\end{figure}

FeSe films with \add{Hall} bar geometry were grown on atomically flat STO (100) substrates with the step-terrace structure using the PLD technique.
The growth conditions are detailed in Ref. \cite{Kobayashi_2022,Kobayashi_2023}.
To prevent degradation of the films by air exposure, amorphous Si with a thickness of approximately 10 nm was deposited at room temperature also using PLD.
The $c$-axis orientation of the grown films was confirmed using X-ray diffraction measurements with Cu K$\alpha$ radiation at room temperature. 
The film thicknesses of the grown films were estimated using FeSe(001) reflections\cite{Kobayashi_2022}. 
The thicknesses of FeSe films in this study were approximately 2.5 and 5 nm, which correspond to 4--5 and 9 layers.
Resistance measurements were performed with the four-terminal method using a physical property measurement system (PPMS).
Thermoelectric measurements were performed with the one-heater-two-thermometer configuration (Fig. \ref{Fig:Fig1}(a)).
The temperature gradient $\nabla T$ was measured with two Cernox thermometers attached to the samples via Cu wires ($\phi=100\  \mathrm{\mu m}$) to reduce the thermal gradient error.
Moreover, phosphor bronze wires were used for voltage measurements.
Because the Si capping layer is insulating, we scratched the contact pads before attaching the wires to the pads to ensure a direct electrical contact with FeSe.
The measurement was performed with fixed temperatures and magnetic fields.
At each measurement, we subtracted background voltage at 
$\nabla T = 0$.
The Nerst signal $N$ in each $B$ was obtained by field antisymmetrization of $e_y=-\frac{V_{xy}/w}{\Delta T/l}$, where $w$ and $l$ are distances between electrodes and thermometers, respectively.
Figure \ref{Fig:Fig1}(b) shows measured voltage $V_{xy}$ as a function of a temperature difference between two thermometers $\Delta T$ at 38 K and 9 T in a 2.5 nm film.
The linear relationship between $\Delta T$ and $V_{xy}$, which is necessary for thermoelectric measurement, was observed.
Validity for the thermoelectric measurement was also confirmed by the measurements in Pt foil and STO bulk (see supplementary material). 
%Figure-----------
%---------------
\section {RESULTS AND DISCUSSION}

\begin{figure}[htpb]
\centering
\includegraphics[width=\linewidth]{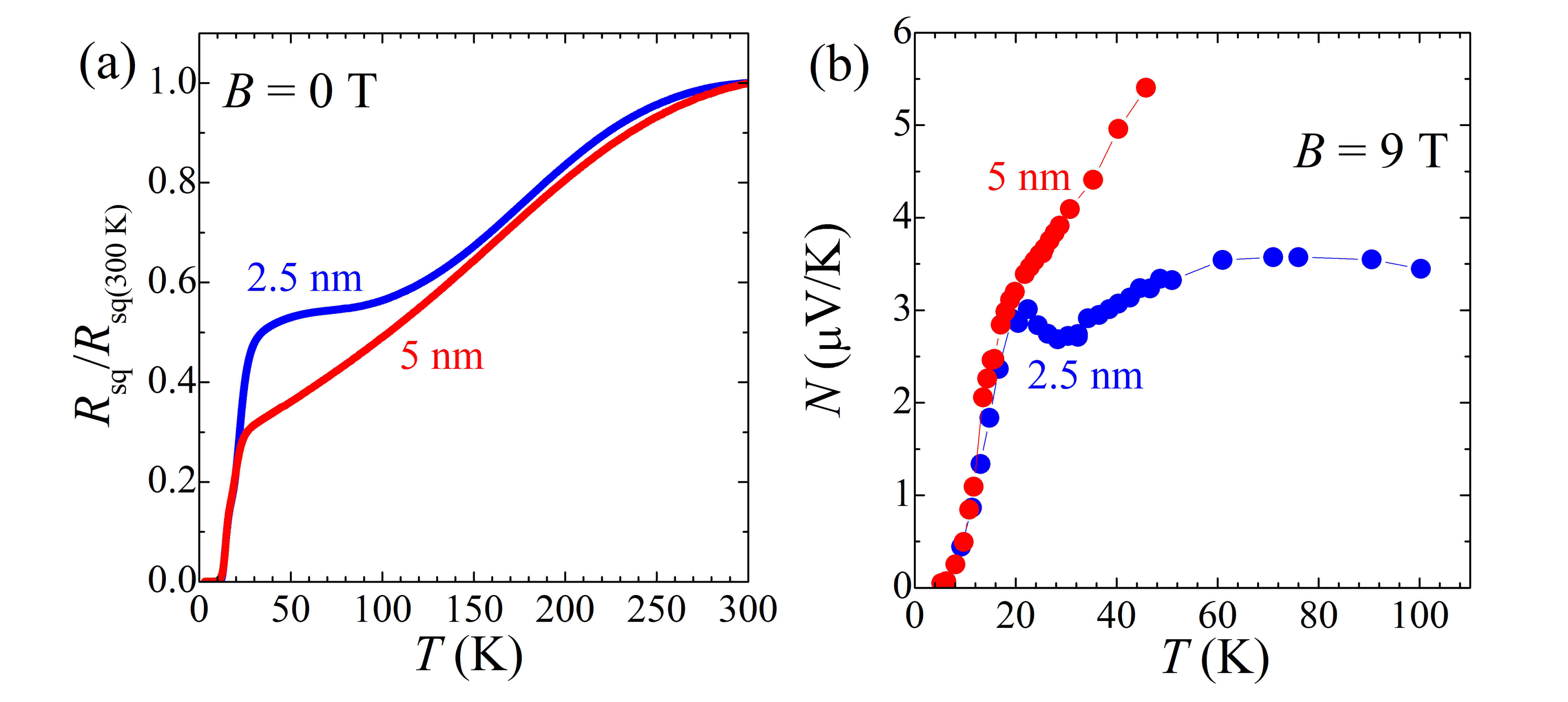}
\caption{\label{Fig:Fig1b} 
(a) Temperature dependences of normalized sheet resistance $R_{sq}$ at 0 T in the grown films.
(b) Nernst signals $N$ as a function of temperature up to 90 K at 9 T in the 2.5 nm film.
}
\end{figure}

Figure \ref{Fig:Fig1b}(a) shows the temperature dependence of normalized sheet resistance $R_{sq}$ for 2.5 and 5 nm films.
These films show metallic behavior and superconducting transitions with $T_{\mathrm{c}}^{\mathrm{onset}} = 27.0$ K, $T_{\mathrm{c}}^{\mathrm{zero}}=8.6$ K for the 2.5 nm film and $T_{\mathrm{c}}^{\mathrm{onset}} = 23.5$ K, $T_{\mathrm{c}}^{\mathrm{zero}}=7.4$ K for {the} 5 nm film.
The temperature dependence of $N$ of both films in 9 T is plotted in Fig. \ref{Fig:Fig1b}(b).
In the 2.5 nm film, $N$ exhibits a broad peak feature at approximately 70 K, followed by a sharper peak at lower temperatures. This observation is discussed later.
%Figure-----------
%---------------
Notably, a similar temperature dependence with the peak feature around 70 K was observed for bulk FeSe\cite{Yang_2017}.
This can be explained as follows: electrons of STO are transferred to a monolayer or a few layers near the FeSe/STO interface as suggested in previous studies\cite{Zhao_2018,Kobayashi_2023}, and the layers distant from the interface retain an electronic structure similar to bulk FeSe.
For the electron-doped layer near the interface, only electron carriers are present, which would lead to a negligible Nernst signal in the normal state due to the Sondheimer cancellation\cite{Sondheimer_1948,Wang_2006,Behnia_2009}.
Indeed, electron-doped FeSe by intercalation of TBA$^+$ does not exhibit the Nernst effect in the normal state\cite{Kang_2020}.
Therefore, the observation of the Nernst effect similar to those in bulk FeSe indicates that Nernst signal in non-doped layers with hole and electron carriers is dominant in the normal state even in the 2.5 nm-thick FeSe/STO.

In 5 nm film, similar to 2.5 nm film, the Nernst effect appeared in the normal state, and $N$ decreased with temperature.
At 45 K, however, the $N$ was larger than that in the 2.5 nm film.
These differences in the magnitude of $N$ can be explained in terms of the difference in the contribution of the non-doped layers.
In a multiband system such as non-doped FeSe, the Nernst signal is sensitive to disorders\cite{Behnia_2009,Oganesyan_2004}.
\add{For instance}, in {semimetallic} Bi, the Nernst signal becomes larger as the residual resistance ratio (RRR) is increased\cite{Galev_1981}.
In the present samples, the 5 nm film exhibited a higher RRR than that of the 2.5 nm film(Fig. \ref{Fig:Fig1b}(a)).
Therefore, the contributions of the non-doped layer in the 5 nm film are considered to be larger than those of the 2.5 nm film.

\begin{figure}[htbp]
%\centering
\includegraphics[width=\linewidth]{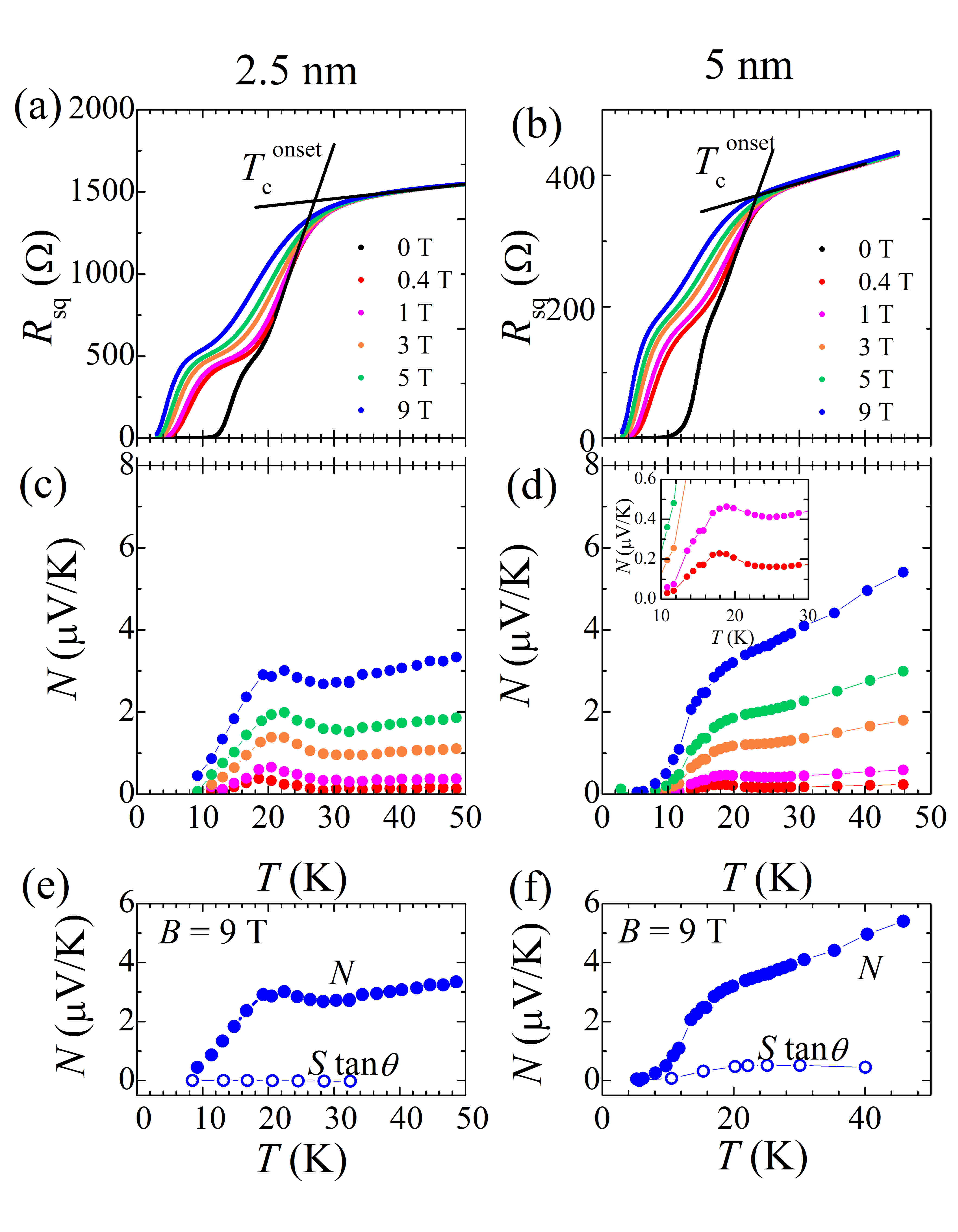}
\caption{\label{Fig:Fig2} 
Temperature dependence of (a) $R_{sq}$ and (c) $N$ at 0--9 T in the 2.5 nm film.
(b) and (d) show the corresponding measurements results obtained for the 5 nm film.
(e) and (f) plot the $N$ and $S\ \mathrm{tan}\theta$ at 9 T for the 2.5 nm and 5 nm films, respectively, for comparison.
}
\end{figure}

Figures \ref{Fig:Fig2}(a) and (c) shows the temperature dependence of $R_{sq}$ and $N$ near \Tconset in 0--9 T for the 2.5 nm film, respectively.
As the magnetic fields increase, the superconducting transition becomes broader.
%, which is similar to two-dimensional superconductors such as cuprate and monolayer films.
In this temperature region, the peak structure of $N$ was observed at all magnetic fields, as mentioned earlier(Fig. \ref{Fig:Fig2}(c)).
The peak structure of Nernst signals corresponding to superconducting transition is generally observed in type-I\hspace{-1.2pt}I superconductors, and this feature is partly attributed to vortex motion induced by thermal force\cite{Behnia_2022}.
The resistive transition broadening caused by the application of magnetic field is attributed to mobile vortices.
Similar phenomenon was also observed in the case of 5 nm film, however, in this case, the peak structure caused by mobile vortices was relatively unclear at $B=9$ T compared with 2.5 nm film(Fig. \ref{Fig:Fig2}(d)).
Because the non-doped layers are expected to be thicker in the 5 nm film than in the 2.5 nm film,
the larger contribution of the non-doped layers obscured the peak structure despite the presence of mobile vortices at 9 T in the 5 nm film.

%Figure-----------
%---------------
%In the presence of the undoped layer 
To discuss the Nernst effect by SC fluctuations, it is important to understand how the non-doped layer affects the measured Nernst signals. 
In general, conductivity tensor $\tilde{\sigma}$ and thermoelectric tensor $\tilde{\alpha}$ follow the relation:
\begin{equation}
    J_e=\tilde{\sigma}\vec{E}-\tilde{\alpha}\nabla T.
    \label{eq:singlelayer}
\end{equation}
Here, $J_e$ is charge current density, and $\vec{E}$ and $\nabla{T}$ are electric field and thermal gradient, respectively.
%For a single layer film, $J_e=0$ because of open circuit conditions. 
In FeSe/STO, a parallel circuit is formed by the non-doped layer and the electron-doped layer near the FeSe/STO interface.
Because Eq. \ref{eq:singlelayer} is valid in each layer, and the total current of both layers becomes zero, the thermoelectric and conductivity tensors in each layer are represented by these equations:
\begin{equation}
\begin{split}
    &j_{e(i)}=\tilde{\sigma}_{(i)}\vec{E}-\tilde{\alpha}_{(i)}\nabla T,\ (i=1,2)\\
    &j_{e(1)}d_1+j_{e(2)}d_2=0.
\end{split} 
\label{eq:twolayer}
\end{equation}
Here, $d_1$ and $d_2$ are the thickness of electron-doped and non-doped layers, respectively.
Note that similar analyses based on the parallel circuit model for the Nernst signal were performed and considered to be reasonable in studies on other materials\cite{Ramos_2013,Lee_2015,Tian_2015}.
From Eq.(\ref{eq:twolayer}), the measured Nernst signals can be derived as
\begin{equation}
    N=-\frac{E_{xy}}{\nabla T}=\frac{{\alpha}_{xy}^{2D}}{{\sigma}_{xx}^{2D}}-\frac{{\alpha}_{xx}^{2D}}{{\sigma}_{xx}^{2D}}\frac{{\sigma}_{xy}^{2D}}{{\sigma}_{xx}^{2D}} 
    \label{eq:N},
\end{equation}
where $\tilde{\alpha}^{2D}=\tilde{\alpha}_{(1)}d_1+\tilde{\alpha}_{(2)}d_2$, and $\tilde{\sigma}^{2D}=\tilde{\sigma}_{(1)}d_1+\tilde{\sigma}_{(2)}d_2$.
Using sheet resistance $R_{sq}$, Seebeck coefficient $S$, and \add{Hall} angle $\theta$, Eq.(\ref{eq:N}) is expressed as
\begin{equation}
    N=R_{sq}\alpha_{xy}^{2D}-S\ \mathrm{tan}\theta.
    \label{eq:N_final}
\end{equation}
In both films, the latter term $S\ \mathrm{tan}\theta$ was very small compared to $N$ as shown in Fig.\ref{Fig:Fig2}(e) and (f).
In this case, $\alpha_{xy}^{2D}$ is calculated using only sheet resistance and Nernst signals\cite{Pourret_2006}.
\add{Because resistance can change the behavior of Nernst signals besides the pure thermoelectric response}, $\alpha_{xy}^{2D}$ was considered more suitable to discuss the Nernst effect by superconductivity originating in the electron-doped layers.

\begin{figure}[htpb]
\centering
\includegraphics[width=\linewidth]{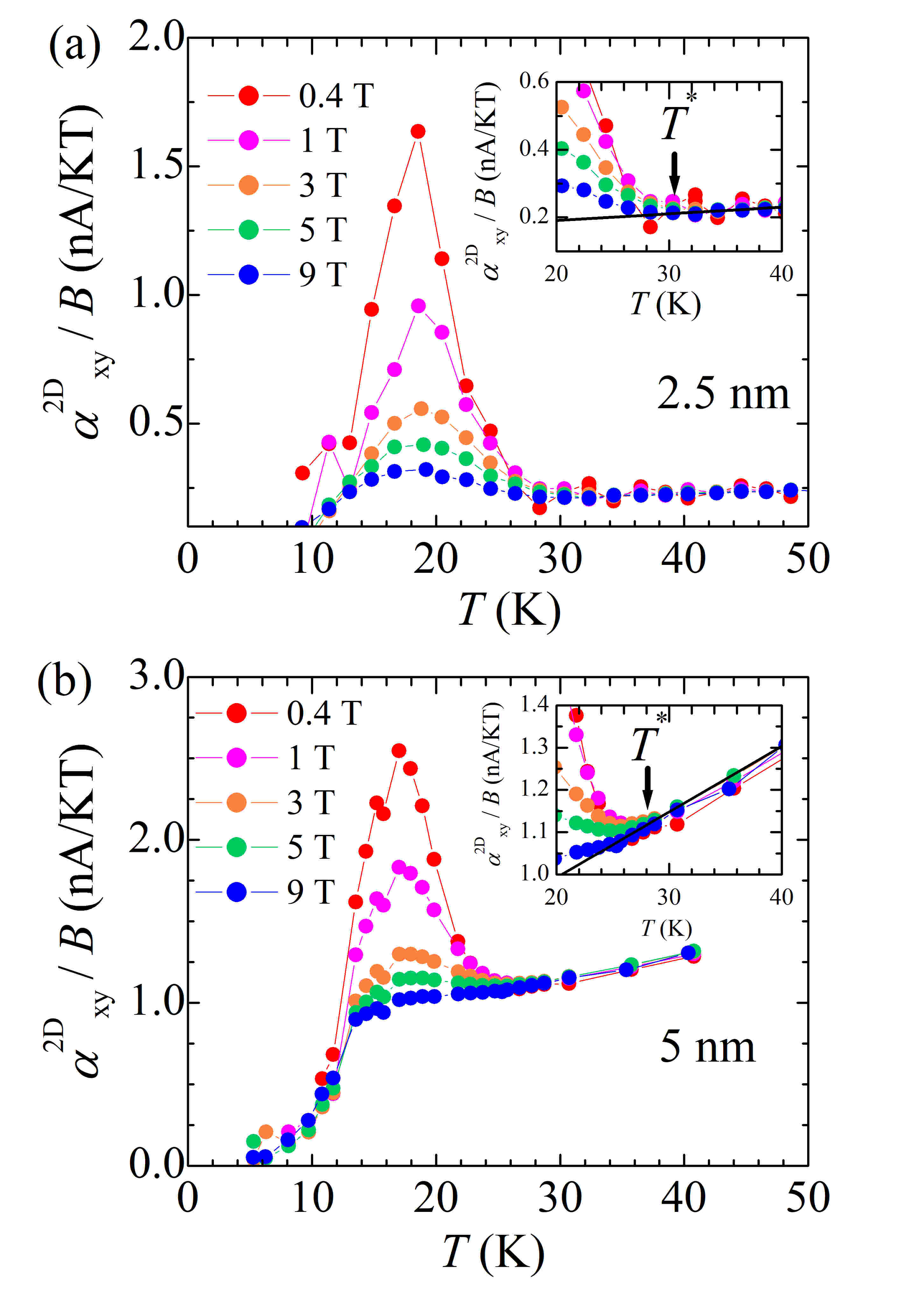}
\caption{\label{Fig:Fig3} (a), (b) Temperature dependence of transverse thermoelectric coefficients divided by magnetic fields $\alpha_{xy}^{2D} /B$ in 0.4, 1, 3, 5, and 9 T of 2.5 nm and 5 nm films, respectively. The insets show the expanded plots near $T^*$.
Solid lines are just the connection between the neighbouring points.}
\end{figure}

The onset of SC fluctuations is usually defined by the temperature where $\alpha_{xy}$ or $N$ starts to increase. 
However, magnetic field dependence is also important to distinguish the Nernst effect by SC fluctuations from that caused by quasi-particles\cite{Cyr-Choini_2018,Jotzu_2023}.
Figure \ref{Fig:Fig3} plots $\alpha_{xy}^{2D}/B$.
As shown, in the case of 2.5 nm film, below $T^* = 30$ K, $\alpha_{xy}/B$ started to increase(Fig. \ref{Fig:Fig3}(a)).
%which suggests additional $\alpha_{xy}$ by SC fluctuations.
%However, the increase of $\alpha_{xy}$ is not necessarily attributed to the SC fluctuations since the other electronic order or changes of electronic structure also affect $\alpha_{xy}^{2D}$ values.
Importantly, while $\alpha_{xy}^{2D}/B$ is independent of $B$ for $T>T^*$, it varies with magnetic fields at lower temperatures.
%The magnetic field dependence of $\alpha_{xy}^{2D}$ differs between those by SC fluctuations or vortex movement and normal carriers.
In the normal state, the thermally diffused normal carrier suffers from the Lorentz force and induces $\alpha_{xy}$ with linear dependence of magnetic fields.
In contrast, $\alpha_{xy}$ by SC fluctuations are rather \add{complicated}, and the magnetic field dependence of $\alpha_{xy}^{2D}$ is non-linear\cite{Cyr-Choini_2018, Jotzu_2023, Ienaga_2024}.
Thus, \add{based on the observed non-linearity, we concluded that the $\alpha_{xy}^{2D}$ just below $T^*$ is originated from SC fluctuations.}
This discussion is valid also for the 5 nm film, in which case, the temperature where $\alpha_{xy}^{2D}$ starts to increase is unclear in $B>3$ T compared to that observed for 2.5 nm film because of the large contribution of the non-doped layer(Fig. \ref{Fig:Fig3}(b)).
However, the variation of magnetic field dependence enables us to determine that $T^*=28$ K for the 5 nm film.
%Figure-----------
%---------------

\begin{figure}[htpb]
\centering
\includegraphics[width=\linewidth]{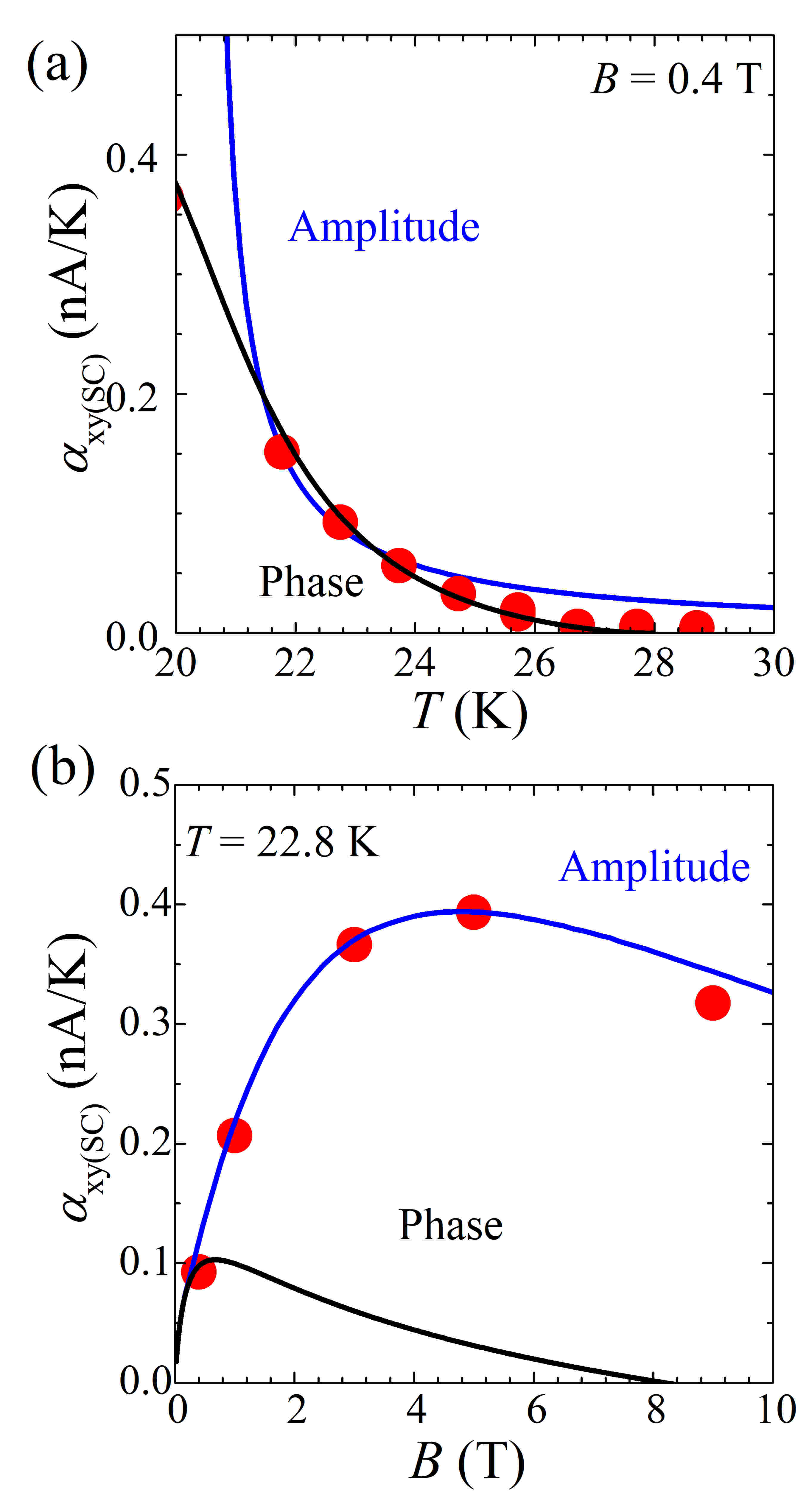}
\caption{\label{Fig:Fig4} Analysis of SC fluctuations for the 5 nm film.
Red points in (a) and (b) denote $\alpha_{xy(\mathrm{SC})}^{2D}$ as a function of temperature at 0.4 T and magnetic field at 22.8 K.
The black curves in (a) and (b) were calculated from Eq. (\ref{eq:fluidmodel}) using parameters in Table (\ref{table:fitpara}).
The blue curve in (a) exhibits the fitting result for Eq.(\ref{eq:gaussian}), while that in (b) is adjusted curve of Ref.\cite{Michaeli_2009} with $B_{\mathrm{peak}}=4.7$ T.}
\end{figure}

As for the origin of $\alpha_{xy}^{2D}$ by SC fluctuations, phase fluctuations of superconducting order parameter $|\Psi|\exp (i\phi)$ is possible due to two-dimensional (2D) nature in ultrathin FeSe/STO.
In a theoretical study, a formula of $\alpha_{xy}^{2D}$ in a 2D superconductor, where Berezinskii–Kosterlitz–Thouless(BKT) transition is expected, was proposed by R. Li and Z. S. She\cite{Li_2017}.
%We compared the $\alpha_{xy}^2D/B$ as a function of temperatures with it calculated based on the XY model which is the standard model for phase fluctuating superconductivity.
In Ref. \cite{Li_2017}, $\alpha_{xy}^{2D}$ based on the phenomenological vortex-fluid model is given by
\begin{equation}
    \alpha_{xy}^{2D}=\frac{\pi \hbar^2}{8m_e\phi_0}\frac{n_s(T)}{T}\frac{B}{B+B_T}\ln\frac{B_{c2}(T)}{B}.
    \label{eq:fluidmodel}
\end{equation}
Here, $m_e$, $n_s(T)$, and $B_{c2}(T)$ are electron mass, microscopic superfluid density, and upper critical field, respectively.
In Ref. \cite{Li_2017}, $B_{c2}$ is considered to weakly depend on temperature based on the reports by Wang \textit{et al}.\cite{Wang_2006}. 
However, the estimation of $B_{c2}$ in Ref. \cite{Wang_2006} is questionable\cite{Chang_2012}.
Although the $B_{c2}(T)$ is usually determined by resistivity data, it is not applicable in this case because resistivity may be insensitive to phase fluctuations\cite{Wang_2006}.
Here, we adopted $B_{c2}(T)=B_{c2}(0)\left(1-(T/T^*)^2\right)/\left(1+(T/T^*)^2\right)$ which is based on Ginzburg-Landau theory.
$B_{c2}(0)$ was determined by resistance data with criterion of 90\%.
In addition, while $n_s(T)=n_s(0)(1-T/T^*)$ in Ref. \cite{Li_2017}, we replaced it with $n_s(T)=n_{s0}(1-(T/T^*)^2)$, which describes $n_s(T)$ of BCS-superconductor near critical temperature\cite{Bert_2012}. 
Because we assumed that Cooper pairs without phase coherence vanish at $T^*$, $n_s(T)$ and $B_{c2}(T)$ become zero at $T^*$.
Note that the modifications in the temperature dependence of $B_{c2}(T)$ and $n_s(T)$ result in only a slight variation in the fitting results.
The effect of thermally induced vortex and antivortex are incorporated in the effective field $B_T=B_l\exp(-2b|T/T_{\mathrm{BKT}}-1|^{-\frac{1}{2}})$, where $B_l$, $T_{\mathrm{BKT}}$, and $b$ are a high-temperature limit of $B_T$, BKT transition temperature, and a constant, respectively.
We assume $B_l=B_{c2}(T_{\mathrm{BKT}})$ following Ref. \cite{Li_2017}.
Furthermore, because $T_{\mathrm{BKT}}$ is not well-defined due to the presence of two-step transitions, we compared the proposed theory and our experimental data by tuning three parameters $b,\ T_{\mathrm{BKT}}$, and $n_{s0}$.
Figure \ref{Fig:Fig4}(a) shows the temperature dependence of $\alpha_{xy}^{2D}$ by SC fluctuations ($\alpha_{xy(\mathrm{SC})}^{2D}$) at 0.4 T in the 5 nm film. These values were obtained by subtraction of linear extrapolation of $\alpha_{xy}^{2D}$ for $T>T^*$.
Eq. (\ref{eq:fluidmodel}) satisfactorily reproduced the data using the parameters in Table \ref{table:fitpara}.
%Black line shows $\alpha_{xy}^{2D}$ by Eq.\ref{eq:fluidmodel} using the parameters in table\ref{table:fitpara}.
%Equation \ref{eq:fluidmodel} was fitted very well for temperature dependence.
However, with the same parameters, the magnetic field dependence of $\alpha_{xy}^{2D}$, as estimated using Eq. (\ref{eq:fluidmodel}), deviates from the experimental data (Fig. \ref{Fig:Fig4}(b)).
$\alpha_{xy(\mathrm{SC})}^{2D}$ shows a peak in $B_{\mathrm{peak}}=$5 T at 22.8 K. 
However, the calculated $\alpha_{xy}^{2D}$ using Eq. \ref{eq:fluidmodel} shows a peak around 1 T, and the maximum value is smaller than peak $\alpha_{xy(\mathrm{SC})}^{2D}$.
Therefore, the observed Nernst effect by SC fluctuations can not be explained by the phase fluctuations scenario. 
\begin{table}
\begin{center}
\caption{Parameters used for plotting the black line in Fig. \ref{Fig:Fig4}.}
  \begin{tabular}{cccccc} \hline 
    $T_c$& $T_\nu$ & $B_{c2}(0)$ & $n_s(0)$ & $B_l$ & $b$ \rule[-5pt]{0pt}{15pt}  \\ \hline     
    18.3 K& 28.0 K & 40.0 T & 2.0$\times 10^{11} $cm$^{-2}$& 40.0 T & $1.06$ \rule[-5pt]{0pt}{15pt}  \\
    \hline
  \label{table:fitpara}
  \end{tabular}
\end{center}
\end{table}
%Thus, the observed fluctuations probably originated from phase fluctuations.
%Figure-----------
%---------------
Another type of SC fluctuations is the amplitude fluctuations of the order parameter.
In 2D superconductor, $\alpha_{xy}^{2D}$ is given as
\begin{equation}
\alpha_{xy}^{2D}\add{/B}=\frac{k_{B}e^2}{6\pi \hbar^2}\frac{\xi_0^2}{1-T/T_{c0}},
\label{eq:gaussian}
\end{equation}
where $\xi_0$ and $T_{c0}$ are a coherence length and mean field transition temperature\cite{Ussishkin_2002}.
In Fig. \ref{Fig:Fig4}(a), blue line shows the curve given by Eq. (\ref{eq:gaussian}) with $\xi_0 = 3.8\ \mathrm{nm}$ and $T_{c0}=20.44\ \mathrm{K}$.
Notably, $\xi_0$ is close to coherence length determined by $\xi=\sqrt{\phi_0/2\pi B_{c2}(0)}$ = 2.86 nm ($B_{c2}(0)=40$ T).
For magnetic field dependence, we employed the theory in Ref.\cite{Michaeli_2009} because Eq.(\ref{eq:gaussian}) is valid only in low-field regimes\cite{Ussishkin_2002}.
\add{This theory involves the diffusion constant $D$ in a superconductor, which is related to coherence length ($D\propto \xi^2$), and coincides with Eq.\ref{eq:gaussian} at low-field limit. It requires  numerical calculations, and only the $B-\alpha_{xy}^{2D}$ curve at $T=1.08\ T_{c0}$ was calculated in Ref. \cite{Michaeli_2009}, which can be used for simulation of  magnetic field dependence\cite{Chang_2012,Tafti_2014}.} \add{Based on the this background, we compared the experimental data near $T=1.08\ T_{c0}$ with the theory in this study.}
Furthermore, the $B_{\mathrm{peak}}$ may be determined by the balance of $\xi(T)$ and magnetic length $l_B=\sqrt{\hbar/(eB)}$\cite{Tafti_2014,Pourret_2009}.
Considering $\xi_0=$ 3.8 nm, $B_{\mathrm{peak}}$ was measured to be 4.7 T at 22.8 K($=1.11\ T_{c0}$), which is close to $1.08\ T_{c0}$.
In Fig. \ref{Fig:Fig4}(b), the adjusted curve in Ref. \cite{Michaeli_2009} with $B_{\mathrm{peak}}$ was plotted as a blue line.
We found that the adjusted curve fitted well in contrast to the phase fluctuation formula(Eq.(\ref{eq:fluidmodel})).
These results suggest that the observed Nernst effect by SC fluctuations originates from amplitude fluctuations.
\add{The absence of clear signatures of phase fluctuations near $T_{\mathrm{c}}^{\mathrm{onset}}$ despite its 2D nature in ultrathin FeSe/STO may be attributed to the fact that BKT-type fluctuations are typically dominant only in the vicinity of the $T_{\mathrm{BKT}}$\cite{Benfatto_2009}. Since $T_{\mathrm{BKT}}$ is expected to be far below $T_{\mathrm{c}}^{\mathrm{onset}}$, the Nernst effect above $T_{\mathrm{c}}^{\mathrm{onset}}$ can be dominated by amplitude fluctuations.}

Finally, combining the diamagnetism measurement, we discuss how superconducting nature emerges \add{in ultrathin FeSe/STO}.
%Figure-----------
\begin{figure}[htpb]
\centering
\includegraphics[width=\linewidth]{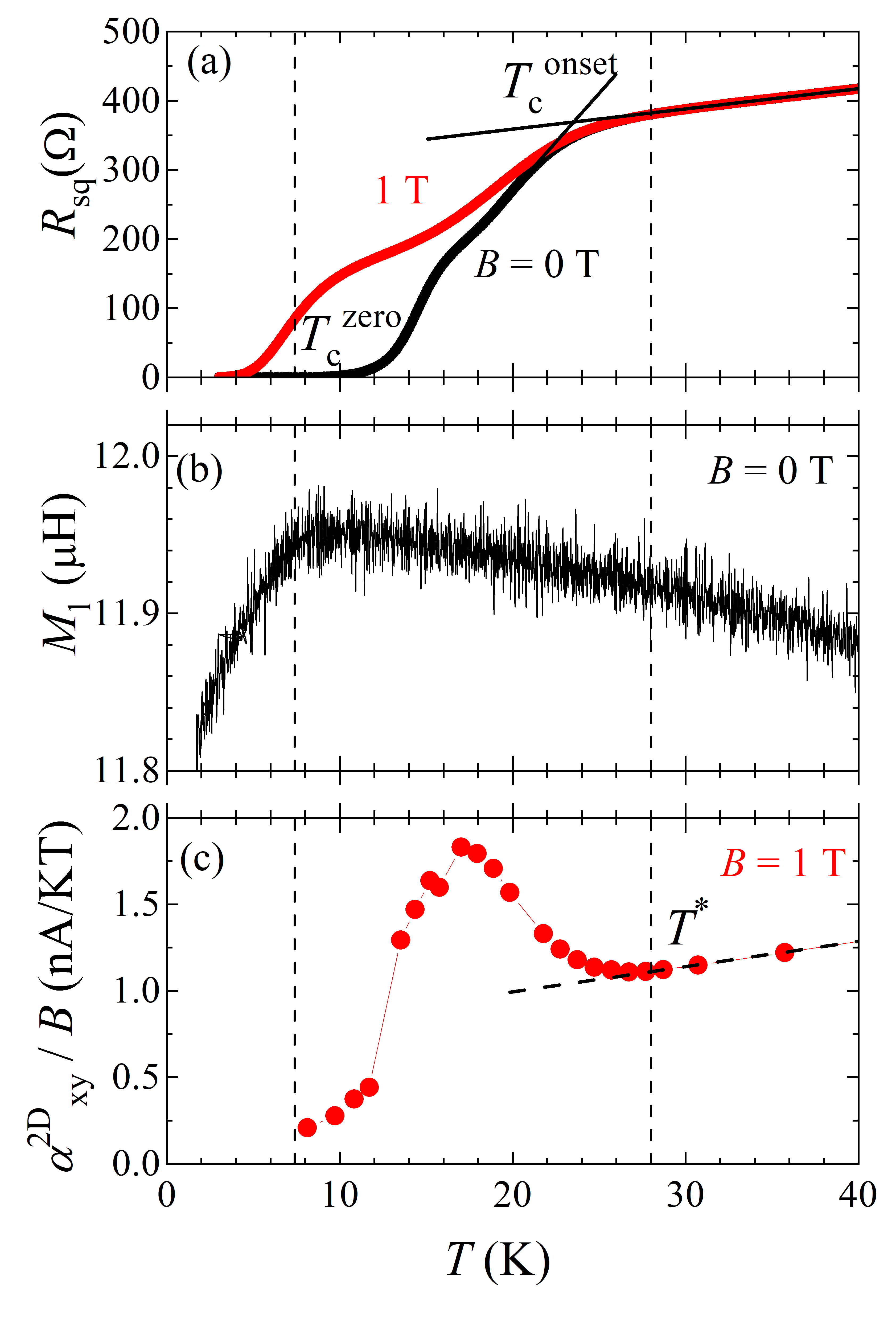}
\caption{\label{Fig:Fig5} Temperature dependence of (a) $R_{sq}$ at 0 and 1 T, (b) real part of mutual inductance $M_1$ in 0 T, and $\alpha_{xy}^{2D}/B$ in 1 T, for the 5 nm film respectively.
}
\end{figure}
In this study, the two-coil mutual inductance experiment was performed for diamagnetism measurement.
Figures \ref{Fig:Fig4}(a), (b), and (c) show temperature dependences of $R_{sq}$ at 0 T and 1 T, the real part of mutual inductance of two coils $M_{1}$ at 0 T, and $\alpha_{xy}^{2D}$ at 1 T, respectively.
Although the decrease of $R_{sq}$ was observed below $T\sim T_{\mathrm{c}}^{\mathrm{onset}}$, the Meissner effect was observed below 8.5 K which is close to \Tczero (= 7.4 K).
This result indicates that the formation of diamagnetic screening is established with a zero-resistance state.
Some previous studies on the Meissner effect measurement reported that the Meissner effect was observed below near \Tczero\cite{Ru_2022,Zhang_2014}, while Feng group reported the Meissner effect below 65 K \cite{Zhang_2015} which is consistent with gap opening temperature reported in ARPES studies\cite{He_2013}. 
Our observed result is consistent with those of the former studies.
Above $T_{\mathrm{c}}^{\mathrm{zero}}$, while the diamagnetic effect was absent, we observed the peak feature in the Nernst effect, which suggests the partial presence of mobile vortices.
Around \Tconset, SC fluctuations should be dominant.
Accordingly, \add{the Nernst effect arising from SC fluctuations was observed up to }$T^*\sim1.2\  T_{\mathrm{c}}^{\mathrm{onset}}$.
This finding is similar to the results observed for iron chalcogenide thin-film superconductor\cite{Nabeshima_2018,Matsumoto_2024,matsumoto_2024_arxiv} and intercalated $\mathrm{(TBA)_xFeSe}$\cite{Kang_2020}.

The absence of the signature of SC fluctuations far above \Tconset is in contrast to the preformed Cooper paring scenario proposed in Ref. \cite{Faeth_2021,Pedersen_2020}.
These studies reported a gap opening, which may be attributed to SC fluctuations, at approximately 70 K which is $1.6\  T_{\mathrm{c}}^{\mathrm{onset}}$.
One possibility for this discrepancy is that the total thickness of the FeSe affects superconducting fluctuations.
%Indeed, the disappearance of the Meissner effect with decreasing film thickness in non-capped film was reported although the reason is not unveiled\cite{Ru_2022}.
Another possibility is that the pseudo-gap phase is not the signature of superconductivity. 
The pseudo gap in cuprates initially attracted attention as a fingerprint of SC fluctuations\cite{Emery_1995,Wang_2006, Lee_2009}. However, many studies reported that other origins are rather likely\cite{Kitano_2006,Kondo_2011,Hashimoto_2015,Cyr-Choini_2018}.
In the case of FeSe/STO, the electronic checkerboard pattern was also reported\cite{Xue_2023}, which can induce electronic reconstruction and the pseudogap.

\section {CONCLUSION}
In conclusion, we investigated the Nernst effect in ultrathin FeSe films on STO.
The temperature dependence of Nernst signals in the normal state is similar to that observed for bulk FeSe, suggesting that the electrons of STO are transferred only to a few layers near the FeSe/STO interface, which is consistent with the findings of previous reports\cite{Zhao_2018,Kobayashi_2023}.
Increase of $\alpha_{xy}^{2D}$ and variation of magnetic field dependence is observed below $T^*\sim 1.2\  T_{\mathrm{c}}^{\mathrm{onset}}$ \add{within our measurement resolution}, which is similar to other Fe-chalcogenide systems.
A comparison to theories suggests that the observed Nenrst effect by the SC fluctuations originates from amplitude fluctuations of the superconducting parameter.
Our obtained results suggest that the origin of the pseudogap in monolayer FeSe/STO may originate in other electronic states rather than superconductivity.
\section {ACKNOWLEDGEMENT}
%\begin{acknowlegement}
We would like to thank S. Okuma at Tokyo Institute of Technology and K. Ienaga at Yamaguchi University for the fruitful discussion. We also thank K. Nakagawa and Y. Shiomi at the University of Tokyo for the discussion of the Nernst experiments. This research was supported by JSPS KAKENHI Grants No. JP24KJ0799 and JP24K06952. 
%\end{acknowlegement}
\bibliography{reference}% Produces the bibliography via BibTeX.

\end{document}